\documentclass[conference]{IEEEtran}

\usepackage{cite} 
\usepackage{amsmath,amssymb,amsfonts}
\usepackage{graphicx}
\usepackage{textcomp}
\usepackage{xcolor}
\usepackage{booktabs} 
\usepackage{hyperref} 
\usepackage{algorithm}
\usepackage{algorithmic}
\usepackage{braket} 

\usepackage{float}
\usepackage{subcaption}
\usepackage{caption}

\title{
Quantum State Continuity Problem (QSCP):\\
A New Primitive for Fork-Resistant Continuity in Quantum-Aware Systems
}

\author{
\IEEEauthorblockN{Samet Ünsal}
\IEEEauthorblockA{
Computer Engineering Student\\
İzmir Katip Çelebi University\\
Email: contact@sametunsal.com
}
}

\begin{document}
\maketitle

\begin{abstract}
We introduce the \emph{Quantum State Continuity Problem (QSCP)}, which asks how a
quantum-aware system can prove that it is a continuation of its earlier execution
under noise, measurement, resets, and adversarial interaction.
Unlike entity authentication or identity proof, QSCP focuses on \emph{process continuity}:
``Is this system the same execution chain as before?''

We formalize QSCP via a security game capturing \emph{fork resistance} and define
a minimal primitive, the \emph{Quantum State Continuity Witness (QSCW)}, which produces
linkable, one-time quantum evidence across time steps.
We specify a round-based
construction (challenge, witness generation, and audit) and propose evaluation
metrics such as audit pass rate and fork-success rate under simple noise models.
\end{abstract}


\section{Introduction}

Long-lived computing systems increasingly rely on remote verification to establish trust.
Device authentication, secure boot, and remote attestation are widely deployed to show that a remote system is authorized and running approved software \cite{attestation}.
These mechanisms typically bind an identity (or device key) to a platform state.

However, identity does not imply \emph{execution continuity}.
Modern environments enable virtualization, snapshotting, rollback, and cloning, allowing an adversary to instantiate multiple divergent executions from a single authenticated past.
This creates a \emph{continuity gap}: classical or post-quantum authentication answers ``does the prover possess a valid secret?'' but not ``is this execution a genuine continuation of a unique past execution?''.
Reset and replay-style adversarial control (e.g., reset attacks) can preserve credential validity while breaking continuity \cite{bellare_reset}.
We argue that continuity should be a first-class security objective, orthogonal to identity.
We call this objective the \emph{Quantum State Continuity Problem (QSCP)}:
can a quantum-aware system produce verifiable evidence that its current execution is a legitimate continuation of its earlier execution under noise, resets, and adversarial interaction?
Quantum information offers physical constraints that are naturally aligned with continuity:
unknown states cannot be cloned (no-cloning) \cite{nocloning,dieks}, and measurement unavoidably disturbs the state.
These constraints are foundational in quantum cryptography \cite{qkd}, but they are usually used in \emph{single-session} settings.
QSCP instead needs evidence that is \emph{time-linked}, so that forked executions cannot both look valid.
In this paper, we introduce a minimal primitive for QSCP, the \emph{Quantum State Continuity Witness (QSCW)}.
QSCW is a stateful witness that evolves across rounds under verifier challenges and produces one-time, linkable quantum evidence.
The central mechanism is \emph{temporal enforcement}: continuity emerges not from a single measurement, but from cumulative constraints across an audit window.

\subsection{Contributions}
\begin{itemize}
    \item We define the \emph{Quantum State Continuity Problem (QSCP)}, formalizing execution continuity as a security goal distinct from identity authentication and attestation \cite{attestation}.
    \item We propose the \emph{Quantum State Continuity Witness (QSCW)} as a minimal primitive that enforces temporal linkage and suppresses forked executions using quantum constraints \cite{nocloning,dieks,helstrom}.
    \item We formalize QSCP via a security game capturing fork-resistance under realistic system adversaries, including reset and snapshot capabilities \cite{bellare_reset}.
    \item We give a round-based toy instantiation and define evaluation metrics: Audit Pass Rate (APR) and Fork Success Rate (FSR).
    \item Through simulation in the NISQ-oriented regime \cite{nisq}, we show that stateless audits do not enforce continuity, whereas temporal enforcement yields approximately exponential suppression of fork success with audit window length.
\end{itemize}

\subsection{Paper Organization}
Section~II reviews background.
Section~III introduces the system and threat model.
Section~IV formalizes QSCP.
Section~V defines QSCW.
Section~VI gives a toy instantiation.
Section~VII discusses security properties and limits.
Section~VIII presents evaluation and results.
Section~IX reviews related work.
Section~X concludes.

\section{Background and Preliminaries}

This section reviews the background concepts required to motivate and
formalize the Quantum State Continuity Problem (QSCP).
We emphasize the distinction between identity authentication and
execution continuity, and discuss why neither classical nor stateless
quantum mechanisms adequately address fork attacks.

\subsection{Identity Authentication vs. Execution Continuity}

Most authentication mechanisms are designed to answer a single question:
whether a prover is authorized to act under a given identity.
This goal is typically achieved through possession of a secret key,
a digital signature, or a hardware-backed attestation report \cite{attestation}.
Execution continuity, however, is a fundamentally different property.
Continuity concerns whether two points in time belong to the same
physical execution, rather than whether they share the same credentials.
A system may successfully authenticate at multiple points in time while
still violating continuity, for example if an adversary duplicates,
replays, or forks an authenticated execution \cite{bellare_reset}.
Classical authentication systems provide no intrinsic mechanism to
distinguish between a single evolving execution and multiple divergent
executions holding identical secrets \cite{attestation}.
As a result, fork attacks remain largely unaddressed at the
authentication layer, despite their relevance in modern deployment
settings such as cloud-based devices and supply-chain verification.

\subsection{Fork Attacks in Modern Systems}

A fork attack occurs when an adversary creates multiple valid future
executions from a single authenticated past.
Such attacks may be enabled by virtualization, snapshotting, rollback,
or cloning of software and hardware state \cite{bellare_reset}.
Importantly, fork attacks do not necessarily violate cryptographic
assumptions \cite{attestation}.
Each fork may correctly authenticate, present valid signatures, and
pass attestation checks.
From the verifier's perspective, the executions are indistinguishable,
even though they represent incompatible physical realities.

\begin{figure}[t]
    \centering
    \includegraphics[width=0.95\linewidth]{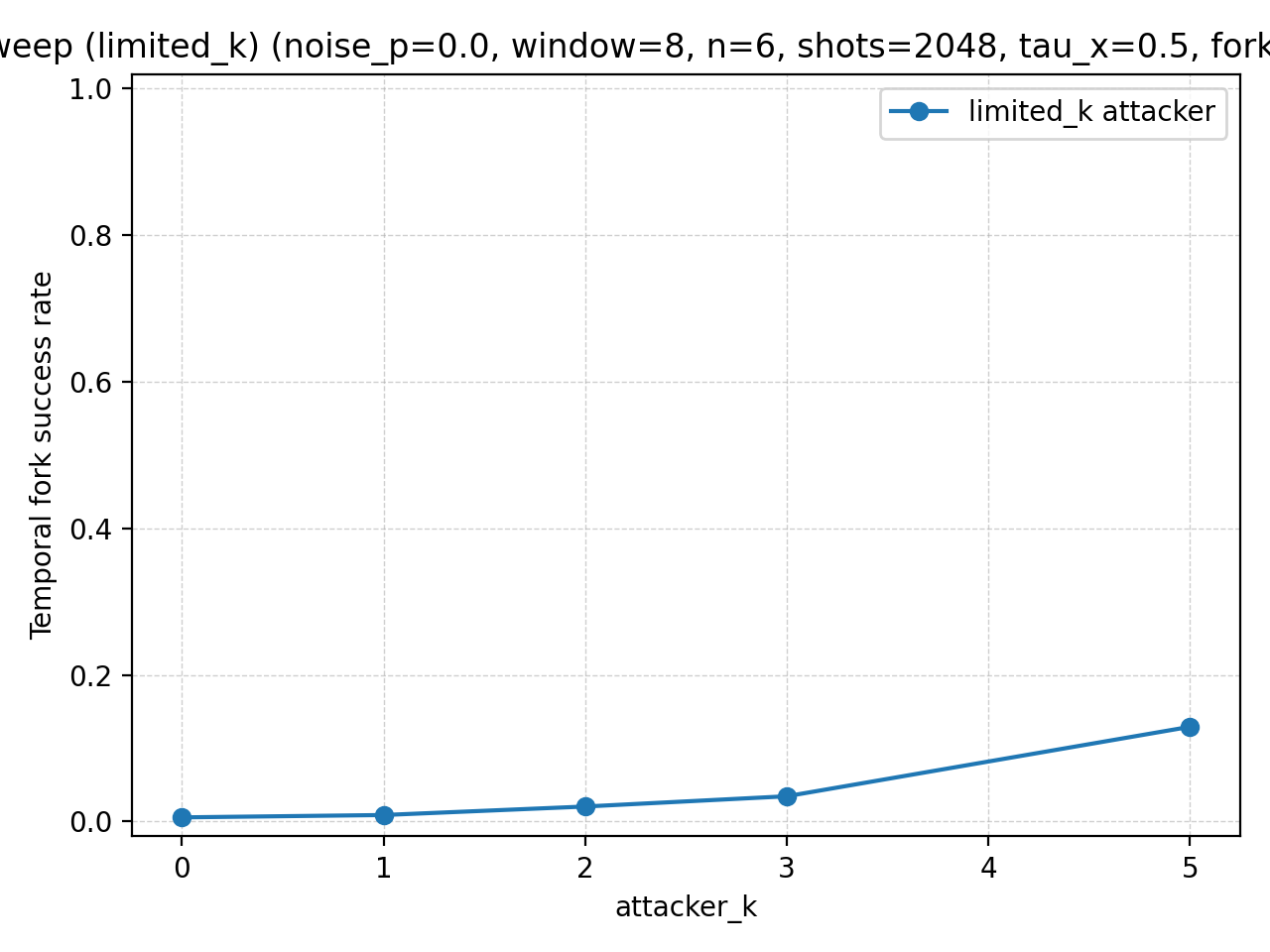}
    \caption{Temporal fork success rate under a limited-$k$ attacker model.
    As attacker capability $k$ increases, fork success rises but remains bounded,
    indicating partial resistance even against adaptive replay strategies.}
    \label{fig:attacker_k_sweep}
\end{figure}

Fork attacks highlight a gap between logical identity and physical
execution \cite{bellare_reset}.
While cryptographic protocols bind actions to keys, they do not bind
actions to a unique temporal evolution.
QSCP aims to formalize and address this gap.

\subsection{Quantum Properties Relevant to Continuity}

Quantum information theory provides physical constraints that have no
classical analogue.
Two properties are particularly relevant to QSCP.

First, the no-cloning theorem prohibits the creation of an identical
copy of an unknown quantum state \cite{nocloning,dieks}.
Second, any measurement that extracts information from a quantum state
necessarily disturbs it \cite{helstrom}.
These properties imply that quantum states cannot be duplicated or
observed without consequence.
In cryptography, these features have been successfully exploited in
quantum key distribution and related protocols \cite{qkd}.
However, such applications typically treat quantum states as ephemeral
resources, consumed within a single protocol round.
They do not enforce continuity across time.

\begin{figure}[t]
    \centering
    \includegraphics[width=0.95\linewidth]{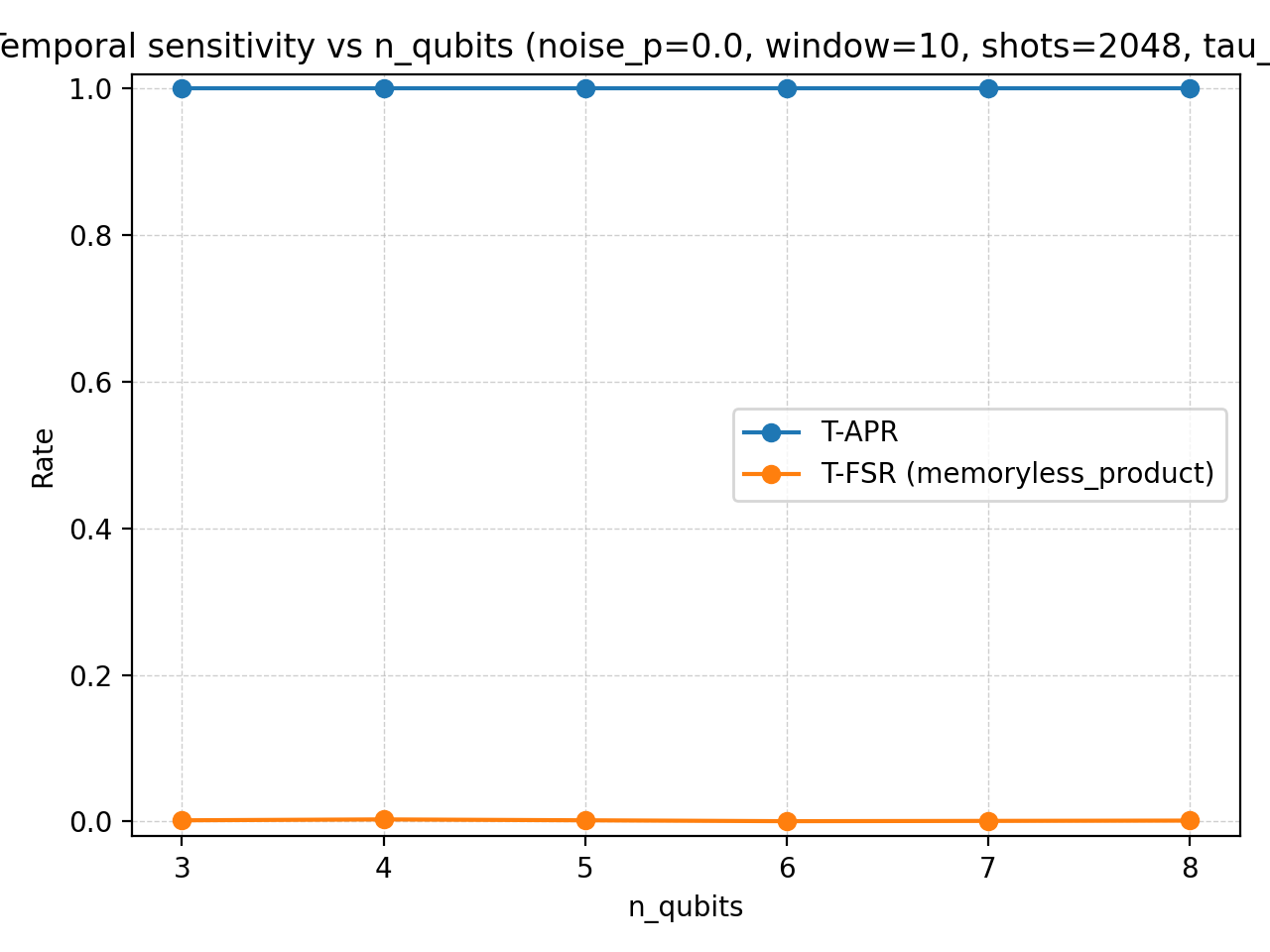}
    \caption{Temporal acceptance and fork sensitivity as a function of the number
    of qubits $n$.
    Temporal acceptance remains high, while stateless fork success
    stays near zero, highlighting robustness independent of system size.}
    \label{fig:n_qubits_sweep}
\end{figure}

\subsection{Limitations of Stateless Quantum Authentication}

\begin{figure}[H]
    \centering
    \includegraphics[width=0.95\linewidth]{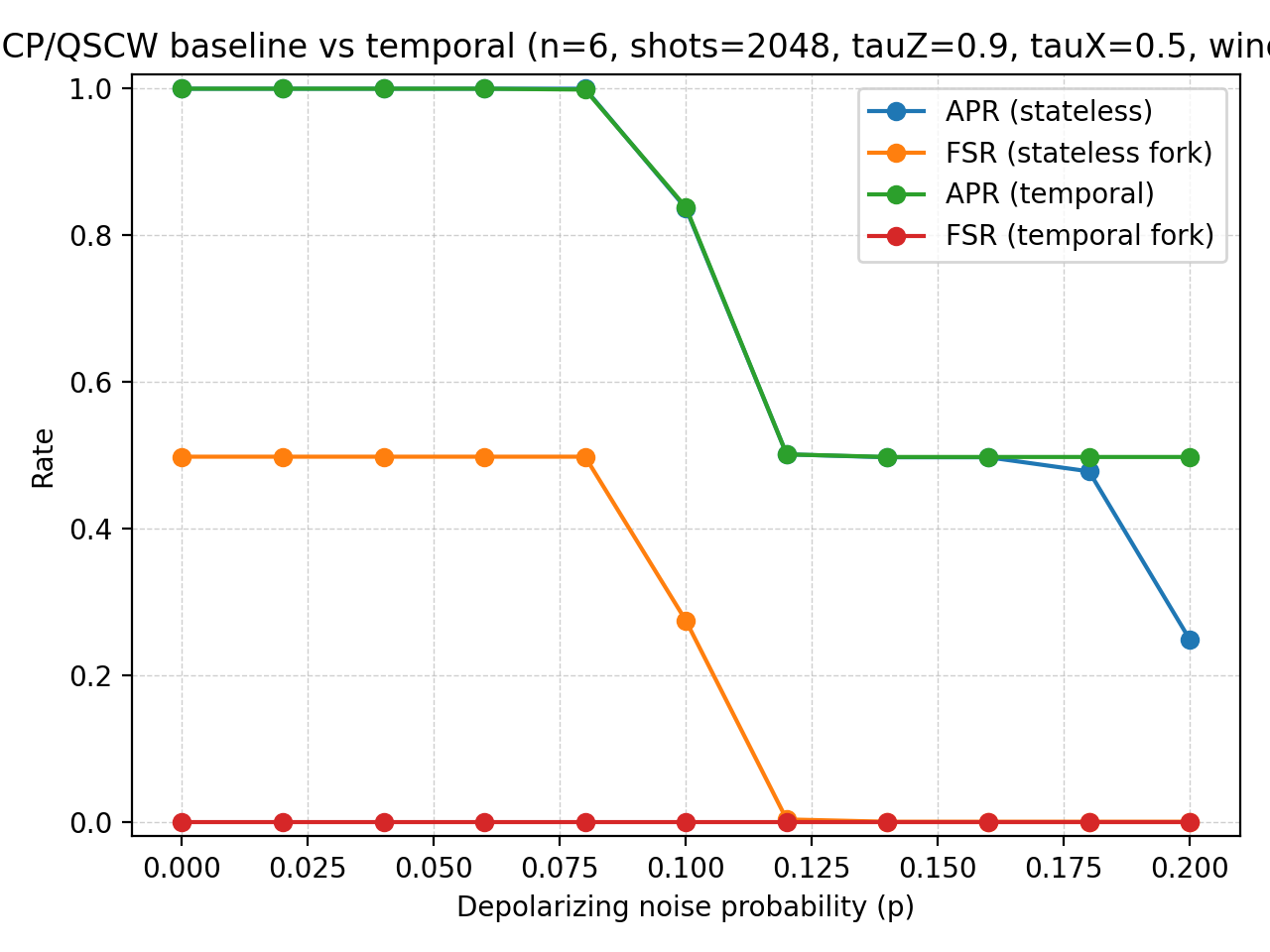}
    \caption{Baseline (stateless) vs temporal construction under depolarizing noise $p$.}
    \label{fig:noise_sweep}
\end{figure}

\begin{figure}[H]
    \centering
    \includegraphics[width=0.95\linewidth]{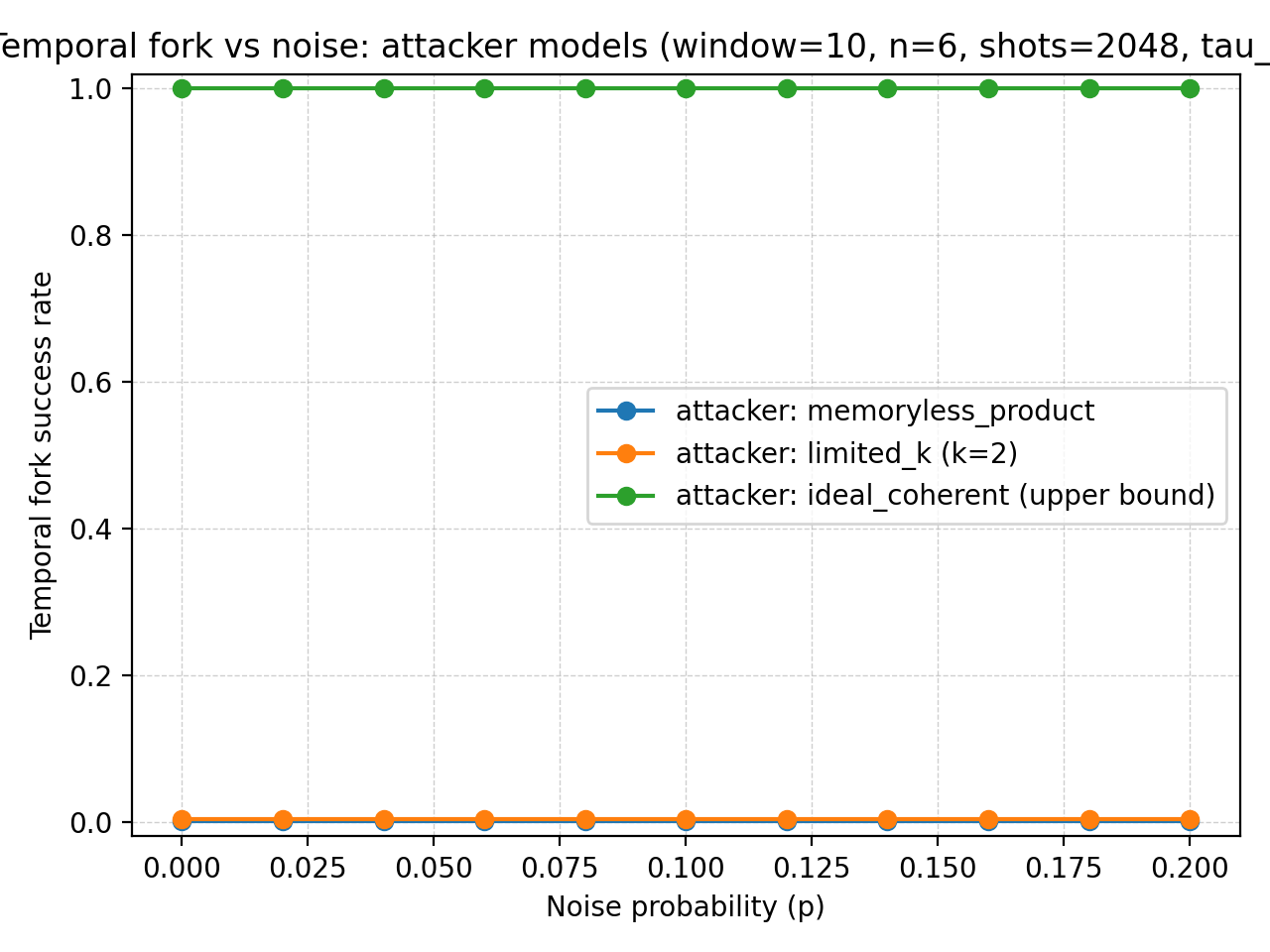}
    \caption{Temporal fork success under different adversary models as a function of noise $p$.}
    \label{fig:noise_sweep_models}
\end{figure}

Several proposals have explored quantum-enhanced authentication
mechanisms, including quantum challenge--response and quantum PUF-like
constructions \cite{damgaard_qid,quantum_puf}.
While these approaches leverage quantum properties, they are typically
stateless: each authentication round is independent of previous rounds.
Stateless quantum authentication inherits the same continuity limitations
as classical schemes \cite{attestation}.
An adversary may measure or approximate a quantum response and reuse it
in a forked execution, succeeding with non-negligible probability
\cite{helstrom}.
As demonstrated in our evaluation, stateless quantum audits achieve
high acceptance for honest behavior but exhibit fork success rates close
to random guessing.
This observation motivates the central thesis of this work: quantum
mechanics alone does not guarantee fork resistance.
Instead, resistance emerges when quantum evidence is forced to evolve
over time, linking present validity to a specific past.~\ref{fig:noise_sweep}~\ref{fig:noise_sweep_models}

\subsection{Towards Temporal Enforcement}

To address QSCP, a mechanism must satisfy two requirements.
First, it must produce evidence that cannot be duplicated without
physical disturbance \cite{nocloning,dieks}.
Second, this evidence must be temporally linked, so that future validity
depends on preserving coherence across multiple rounds \cite{nisq}.
The Quantum State Continuity Witness introduced in this work satisfies
these requirements by design.
By embedding verifier challenges into the temporal evolution of a
quantum state, QSCW enforces execution continuity and suppresses fork
attacks through cumulative auditing \cite{ghz,nisq}.

\begin{figure}[t]
    \centering
    \includegraphics[width=0.95\linewidth]{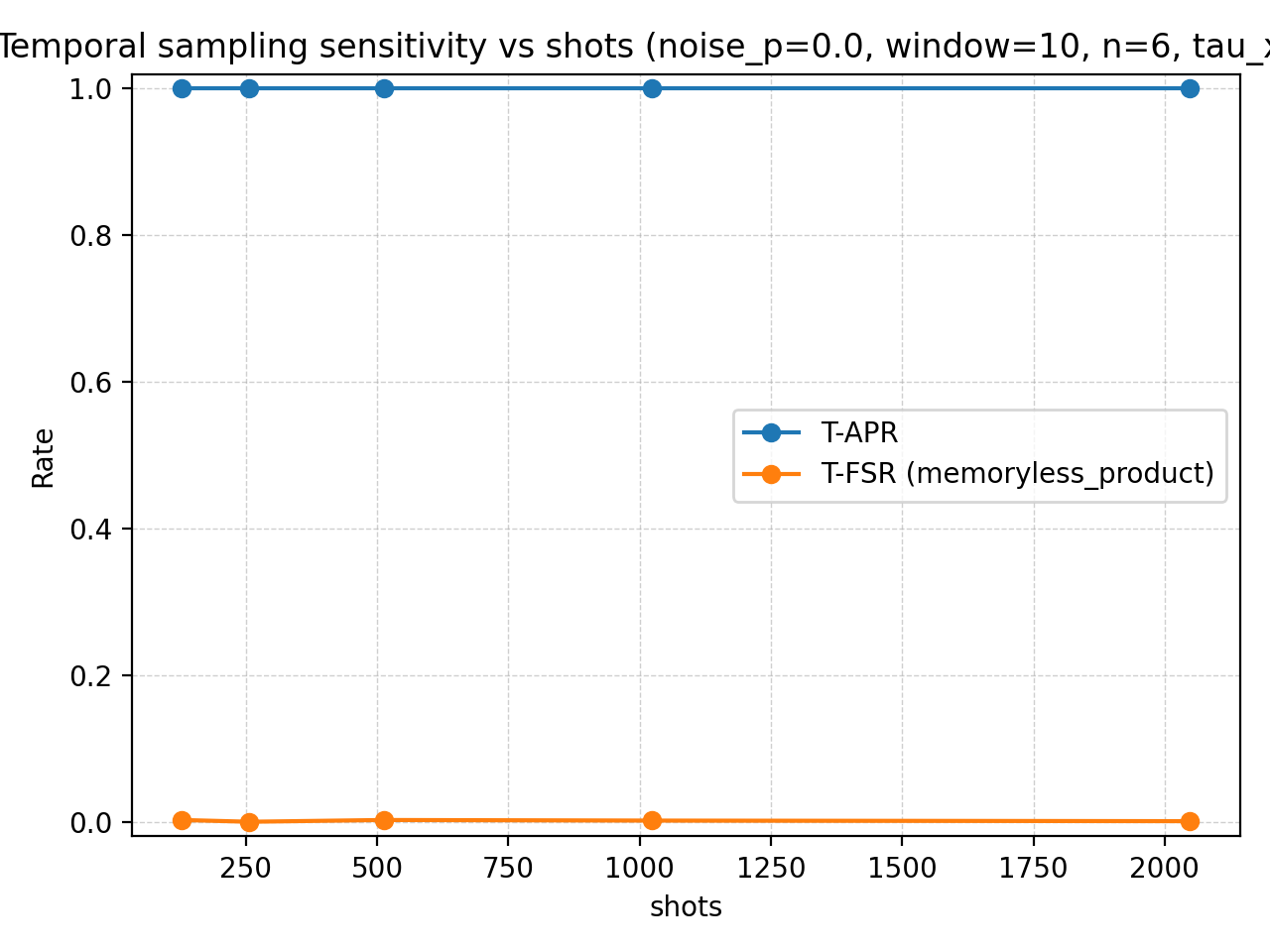}
    \caption{Temporal audit robustness with respect to the number of measurement shots.
    Acceptance remains stable while fork success stays negligible, indicating
    resilience to finite sampling effects.}
    \label{fig:shots_sweep}
\end{figure}

\begin{figure}[t]
    \centering
    \includegraphics[width=0.95\linewidth]{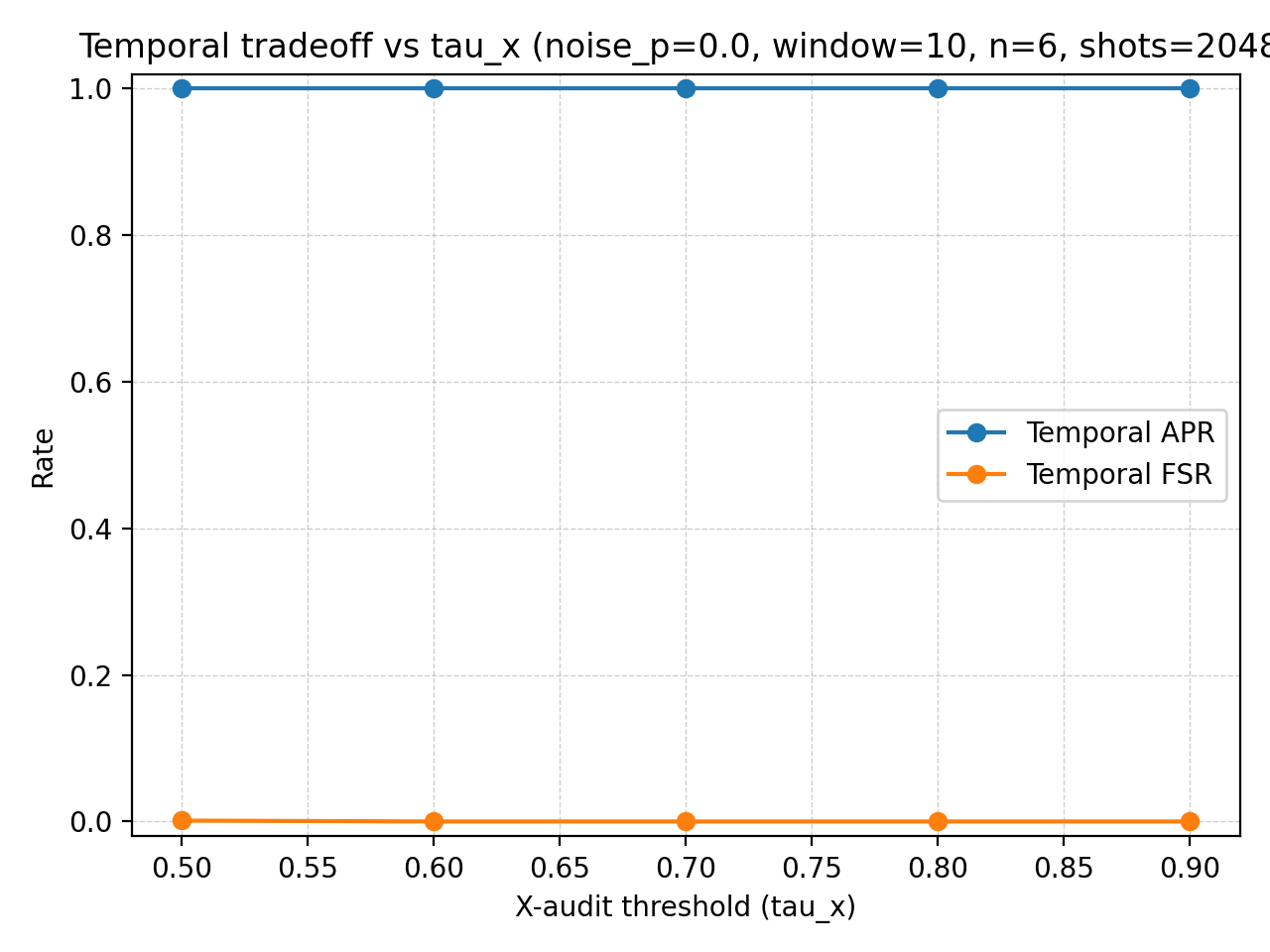}
    \caption{Tradeoff between temporal acceptance and fork suppression as a function
    of the audit threshold $\tau_x$.
    Temporal protocols maintain strong fork
    resistance across a wide threshold range.}
    \label{fig:tau_x_sweep}
\end{figure}

\begin{figure}[t]
    \centering
    \includegraphics[width=0.95\linewidth]{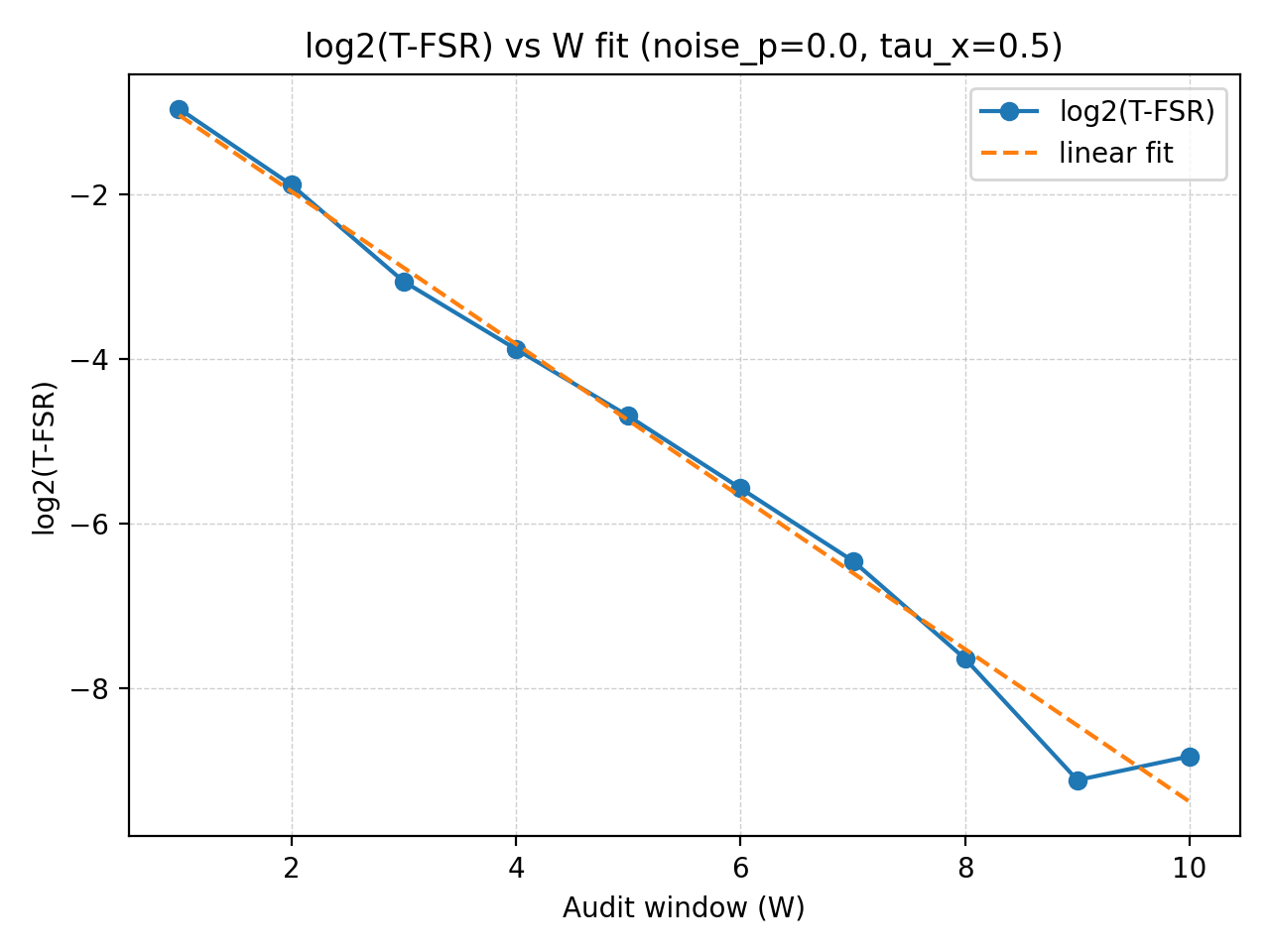}
    \caption{Exponential decay of temporal fork success with audit window size $W$.
    The linear behavior in $\log_2(\mathrm{T\text{-}FSR})$ confirms an approximately
    $2^{-W}$ suppression law.}
    \label{fig:window_decay_fit}
\end{figure}

\begin{figure}[t]
    \centering
    \includegraphics[width=0.95\linewidth]{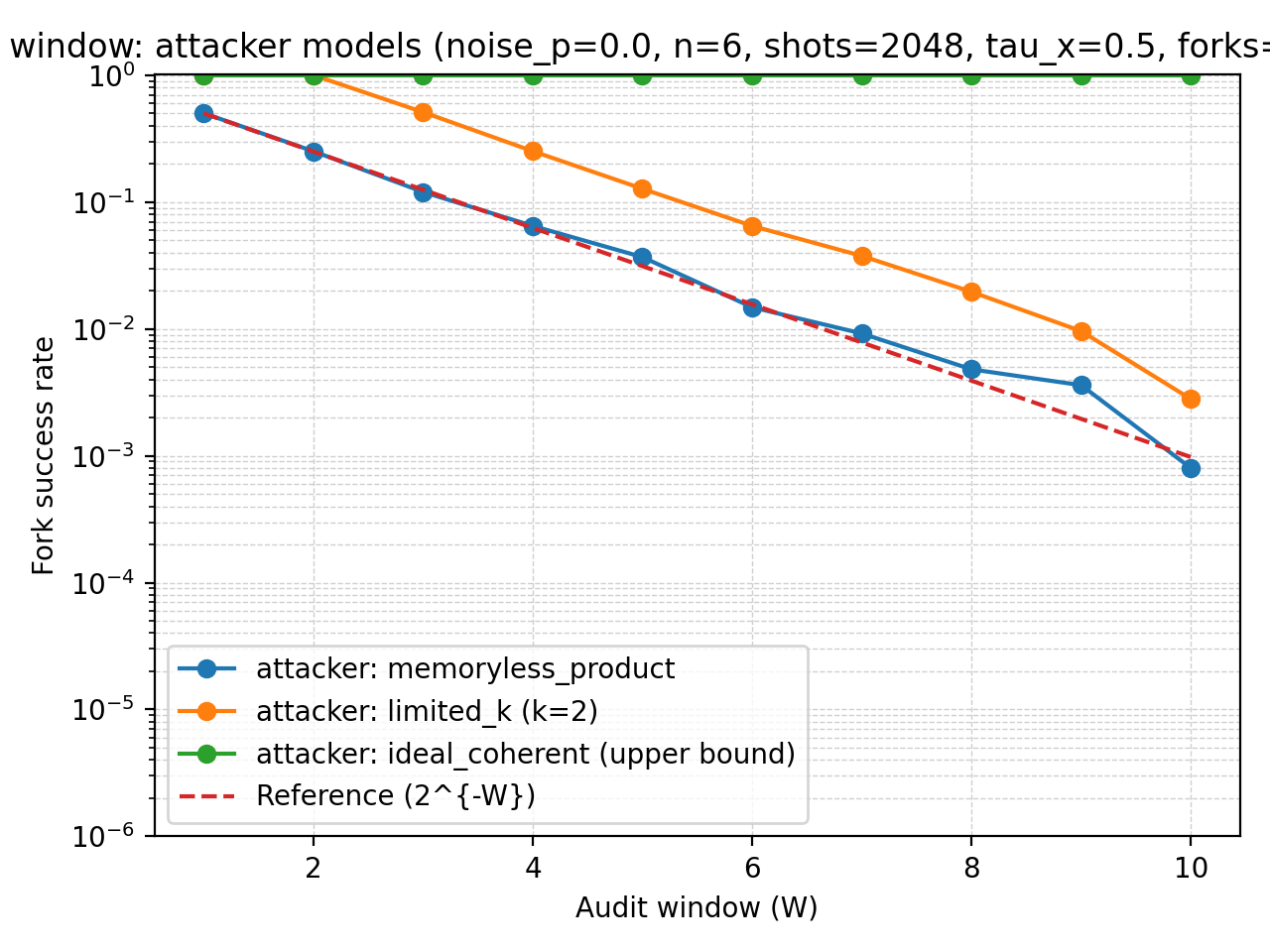}
    \caption{Fork success as a function of audit window $W$ for different attacker
    models.
    Empirical results closely follow the theoretical $2^{-W}$ reference.}
    \label{fig:window_sweep_models}
\end{figure}

\section{System and Threat Model}

This section describes the system setting considered in this work and
formalizes the adversarial capabilities relevant to the Quantum State
Continuity Problem (QSCP).
Our model is intentionally minimal and system-oriented, focusing on
execution continuity rather than full authentication protocols
\cite{attestation,bellare_reset}.

\subsection{System Model}

We consider an interactive system consisting of a prover device and a
verifier.
The prover maintains an internal state and participates in a sequence
of rounds indexed by discrete time steps $t = 1, \dots, T$.
At each round, the verifier issues a fresh challenge and may request
quantum evidence from the prover.
The prover responds by producing a quantum witness derived from its
current internal state.
The verifier performs an audit on the received witness and records the
outcome.
A final decision regarding execution continuity is made after observing
the outcomes of multiple rounds.
The system operates alongside conventional authentication mechanisms,
such as cryptographic keys, secure boot, or remote attestation
\cite{attestation}.
QSCP does not assume that these mechanisms are absent or broken.
Instead, it addresses a complementary security goal: enforcing that a
single authenticated execution cannot be forked into multiple valid
future executions \cite{bellare_reset}.

\subsection{Temporal Model}

Time is modeled as a sequence of logical rounds rather than physical
clock cycles.
The prover is not required to maintain strict real-time guarantees,
only that each round corresponds to a distinct interaction with the
verifier.
Between rounds, the prover may experience noise, resets of classical
state, or environmental disturbances.
Quantum state evolution is assumed to be imperfect and subject to noise,
reflecting near-term quantum devices \cite{nisq}.
QSCP tolerates such imperfections, provided that the execution remains
a single continuous evolution.

\subsection{Adversarial Capabilities}

The adversary is assumed to have full control over the prover’s
classical environment.
This includes the ability to observe protocol transcripts, reset
classical memory, snapshot software state, and adaptively choose future
actions based on prior interactions \cite{bellare_reset}.
Crucially, the adversary may attempt to \emph{fork} the execution by
creating multiple divergent future states from a single authenticated
past.
Each fork interacts independently with the verifier and attempts to
appear as a valid continuation of the same execution history.
The adversary is not assumed to break cryptographic primitives or
violate physical laws.
In particular, the adversary cannot clone unknown quantum states and
cannot extract information from a quantum state without causing
disturbance \cite{nocloning,dieks,helstrom}.
These constraints reflect standard assumptions in quantum information
theory.

\subsection{Adversary Models}

To capture realistic system threats, we consider several adversary
classes.
A \emph{memoryless adversary} replaces the quantum witness with a
classical or separable approximation after each round
\cite{helstrom}.
A \emph{limited-memory adversary} retains partial information about past
challenges but cannot preserve full quantum coherence over time.
Finally, we consider an \emph{ideal coherent adversary} that can maintain
perfect quantum coherence across rounds \cite{nocloning}.
The ideal coherent adversary represents an upper bound rather than a
realistic threat model.
QSCP does not claim security against such an adversary.
Instead, our goal is to characterize the regime in which execution
continuity can be enforced against realistic system-level attackers.

\subsection{Success Conditions}

The adversary succeeds if it can produce two or more forked executions
that all pass the verifier’s audits as valid continuations of the same
past execution \cite{bellare_reset}.
The defender succeeds if at least one fork is detected and rejected
within a bounded number of rounds.
We emphasize that QSCP does not aim to identify or authenticate a
specific entity.
Its objective is solely to enforce continuity: ensuring that a valid
future execution is uniquely linked to a single past.

\section{The Quantum State Continuity Problem}

This section formalizes the Quantum State Continuity Problem (QSCP).
QSCP captures the security objective of enforcing that a system’s current
execution is a legitimate continuation of a unique past execution, even
in the presence of adversarial interaction \cite{bellare_reset}.
The problem is intentionally defined independently of specific
cryptographic protocols or hardware assumptions \cite{attestation}.

\subsection{Problem Definition}

Informally, QSCP asks whether a system can produce evidence that binds
its present behavior to a single past execution history.
Unlike identity authentication, which binds actions to an entity or key,
QSCP binds actions to a temporal evolution \cite{bellare_reset}.
Formally, consider an interactive process between a prover and a
verifier over a sequence of rounds $t = 1, \dots, T$.
At each round, the prover produces evidence derived from an internal
state that may evolve over time.
An execution is said to be \emph{continuous} if there exists a single
underlying state trajectory that generates all accepted evidence.
A \emph{fork} occurs if an adversary can generate two or more divergent
state trajectories from a common past, such that each trajectory
produces evidence accepted by the verifier as a valid continuation
\cite{bellare_reset}.
The goal of QSCP is to ensure that the probability of such forked
acceptance is negligible, while preserving high acceptance for honest,
non-forked executions \cite{helstrom}.

\subsection{Security Game}

We define QSCP using a game-based formulation \cite{goldreich}.

\begin{algorithm}[t]
\caption{QSCP Security Game ($\mathcal{G}_{\text{QSCP}}$)}
\label{alg:security_game}
\begin{algorithmic}[1]
\REQUIRE Security parameter $\lambda$, Audit window $W$
\STATE \textbf{Setup:} Challenger $\mathcal{C}$ initializes Prover $P$ with secret state.
\STATE \textbf{Phase 1 (History):} Adversary $\mathcal{A}$ interacts with $P$ for $t=1 \dots T_{fork}$.
\STATE \textbf{Forking:} At $T_{fork}$, $\mathcal{A}$ creates two branches: $B_0$ (original) and $B_1$ (forked/cloned).
\STATE \textbf{Phase 2 (Challenge):} 
\FOR{$i=1$ to $W$}
    \STATE $\mathcal{C}$ sends random challenge $c_i$ to both $B_0$ and $B_1$.
    \STATE $\mathcal{A}$ produces evidences $E^{(0)}_i$ and $E^{(1)}_i$.
    \STATE $\mathcal{C}$ audits both evidences.
\ENDFOR
\STATE \textbf{Decision:} 
\IF{Verifier accepts BOTH $B_0$ and $B_1$ for all steps in $W$}
    \RETURN $\mathcal{A}$ wins (1)
\ELSE
    \RETURN $\mathcal{A}$ loses (0)
\ENDIF
\end{algorithmic}
\end{algorithm}

The game proceeds as follows:
\begin{enumerate}
    \item The challenger initializes the prover state and interacts with
    it honestly for a prefix of rounds.
    \item At a time of the adversary’s choosing, the adversary may fork
    the execution into multiple branches \cite{bellare_reset}.
    \item Each branch continues interacting independently with the
    verifier, receiving fresh challenges and producing evidence.
    \item The verifier audits each branch over a bounded number of rounds
    and records the outcomes.
\end{enumerate}

The adversary wins if all forked branches are accepted as valid
continuations of the same past execution.
The defender wins if at least one branch is rejected within the audit
window.
QSCP is satisfied if the adversary’s winning probability is sufficiently
small under realistic adversarial constraints \cite{goldreich}.

\subsection{Correctness and Completeness}

A QSCP construction must satisfy correctness for honest executions.
Specifically, an honest prover that follows the prescribed evolution
should be accepted with high probability, despite noise and imperfect
quantum operations \cite{nisq}.
Correctness ensures that enforcing continuity does not unduly penalize
legitimate executions.
This property is essential for practical deployment, where noise and
minor disturbances are unavoidable \cite{nisq}.

\subsection{Fork Resistance}

Fork resistance is the central security requirement of QSCP.
Intuitively, fork resistance demands that an adversary cannot create
multiple valid futures from a single authenticated past \cite{bellare_reset}.
We emphasize that fork resistance is a temporal property.
A single successful audit does not imply continuity.
Instead, continuity must be enforced cumulatively across multiple rounds,
so that any disruption caused by measurement, approximation, or
classicalization is eventually detected \cite{nocloning,helstrom}.

\subsection{Limitations}

QSCP does not aim to provide unconditional security against all possible
adversaries.
In particular, an adversary capable of maintaining perfect quantum
coherence across time and branches may defeat continuity enforcement
\cite{nocloning,dieks}.
Such an adversary represents an idealized upper bound rather than a
realistic system threat.
QSCP is therefore best understood as a framework for characterizing the
trade-offs between temporal enforcement, adversarial capability, and
practical feasibility \cite{nisq}.

\section{Quantum State Continuity Witness}

This section introduces the \emph{Quantum State Continuity Witness (QSCW)},
a minimal security primitive designed to address the Quantum State
Continuity Problem (QSCP) \cite{bellare_reset}.
QSCW is intentionally defined independently of specific protocols or
hardware realizations, focusing instead on the properties required to
enforce execution continuity \cite{attestation}.

\subsection{Definition}

A Quantum State Continuity Witness is a stateful mechanism that produces
quantum evidence over a sequence of rounds.
At each round $t$, the witness generates evidence $W_t$ that depends on
both the verifier’s challenge and the witness’s prior internal state
\cite{goldreich}.
The defining feature of QSCW is \emph{temporal dependency}:
valid evidence at time $t$ cannot be generated without preserving the
evolution of the witness through all preceding rounds.
Any attempt to duplicate, replay, or branch the witness state disrupts
this dependency and is detectable through subsequent audits
\cite{nocloning,helstrom}.
Formally, a QSCW consists of:
\begin{itemize}
    \item an initialization procedure that prepares an initial quantum
    state,
    \item an update rule that evolves the state in response to verifier
    challenges, and
    \item an evidence generation procedure that produces audit material
    from the current state.
\end{itemize}

\subsection{Statefulness and Continuity}

Unlike stateless authentication mechanisms, QSCW is explicitly stateful.
The witness state is not reset between rounds;
instead, it accumulates
information derived from the interaction history \cite{bellare_reset}.

Statefulness is not an implementation artifact but a security
requirement.
If the witness were reset or reinitialized at each round, the adversary
could replay or approximate valid responses independently for each
audit, enabling fork attacks \cite{helstrom}.
By enforcing stateful evolution, QSCW ties future validity to past
behavior.

\subsection{Security Properties}

A QSCW construction is expected to satisfy three core properties.
\paragraph{Completeness}
An honest execution that preserves a single continuous witness state
should be accepted with high probability across all rounds, despite
noise and imperfect quantum operations \cite{nisq}.
\paragraph{Fork Resistance}
An adversary attempting to fork the execution into multiple branches
should fail with high probability.
Specifically, the probability that all forked branches are accepted as
valid continuations should decrease as the audit window increases
\cite{bellare_reset}.
\paragraph{Replay Resistance}
Evidence produced at a given round should not be reusable in a different
temporal context.
Any attempt to replay or reuse quantum evidence outside its intended
round should result in detectable inconsistencies \cite{nocloning}.

\subsection{Temporal Enforcement Mechanism}

\begin{algorithm}[H]
\caption{Temporal QSCW Round Procedure (High-Level)}
\label{alg:temporal_qscw}
\begin{algorithmic}[1]
\STATE \textbf{Inputs:} round index $t$, witness state $\rho_{t-1}$, verifier challenge $c_t$, thresholds $(\tau_x,\tau_z)$
\STATE \textbf{Output:} updated witness $\rho_t$, evidence transcript $E_t$, accept/reject
\STATE Verifier samples a random challenge $c_t$ and sends to prover.
\STATE Prover applies history-dependent update $U(c_t)$ on $(\rho_{t-1} \otimes \ket{ancilla})$ to obtain $\rho_t$.
\STATE Prover generates evidence $E_t$ by measuring only audit registers / ancilla (not the full data state).
\STATE Verifier computes audit statistics from $E_t$ and compares to $(\tau_x,\tau_z)$.
\STATE \textbf{Accept} if audit passes; otherwise \textbf{Reject}.
\end{algorithmic}
\end{algorithm}

QSCW enforces continuity through cumulative auditing \cite{helstrom}.
Rather than relying on a single decisive measurement, the verifier
collects audit outcomes across multiple rounds.
Each audit constrains the set of possible witness states compatible with
the observed evidence.
As audits accumulate, the space of states consistent with multiple
divergent futures rapidly collapses.
An adversary that disrupts the witness through measurement or
approximation may succeed in isolated rounds but will eventually be
detected as constraints compound over time \cite{nocloning,dieks}.

\subsection{Scope and Non-Goals}

QSCW is not intended to replace existing authentication or attestation
mechanisms \cite{attestation}.
It does not identify entities, manage keys, or establish secure channels.
Instead, it provides a complementary capability: enforcing that a valid
future execution is uniquely linked to a single past execution
\cite{bellare_reset}.
We emphasize that QSCW does not guarantee security against idealized
adversaries capable of maintaining perfect quantum coherence across time
and branches \cite{nocloning,dieks}.
Such adversaries define an upper bound on achievable security rather than
a realistic deployment scenario.
QSCW should therefore be understood as a building block for
quantum-aware security systems, rather than a complete solution
\cite{nisq}.

\section{Example Instantiation}

To demonstrate the feasibility of the Quantum State Continuity Witness
primitive, we present a simple toy instantiation.
The goal of this construction is not optimal security or efficiency,
but to illustrate how temporal enforcement can suppress fork attacks in
practice \cite{bellare_reset,nisq}.

\subsection{Witness Initialization}

The witness is initialized as an entangled multi-qubit state.
Specifically, we consider an $n$-qubit GHZ state of the form
\[
\ket{\mathrm{GHZ}} = \frac{1}{\sqrt{2}}\left(\ket{0}^{\otimes n} +
\ket{1}^{\otimes n}\right).
\]
This state exhibits strong correlations under complementary
measurements and serves as a convenient vehicle for encoding temporal
information \cite{ghz}.
The initial GHZ state is prepared once at the beginning of the
execution.
Importantly, the witness is not reinitialized between rounds
\cite{bellare_reset}.

\subsection{Challenge-Dependent Evolution}

At each round, the verifier issues a fresh classical challenge.
The parity of the challenge bits determines an update applied to the
witness state.
Concretely, if the challenge parity is odd, a phase-flip operation is
applied to one qubit of the GHZ state.
This transforms the witness between the $\mathrm{GHZ}^+$ and
$\mathrm{GHZ}^-$ variants, encoding temporal information into the global
phase of the state \cite{ghz}.
If the challenge parity is even, no update is applied.
This update rule ensures that the witness state evolves deterministically
as a function of the interaction history.
Valid future evidence therefore depends on preserving the accumulated
phase information across rounds \cite{nocloning}.

\subsection{Audit Procedure}

At each round, the verifier performs an audit by requesting a
measurement of the witness in a randomly chosen basis.
Two complementary bases are considered \cite{helstrom}.

In the computational ($Z$) basis, both $\mathrm{GHZ}^+$ and
$\mathrm{GHZ}^-$ yield identical outcome distributions, providing no
information about the encoded phase.
In the complementary ($X$) basis, the two variants differ in their parity
statistics: one favors even parity outcomes, while the other favors odd
parity outcomes \cite{ghz,helstrom}.
The verifier accepts or rejects the audit outcome based on whether the
observed statistics are consistent with the expected parity implied by
the accumulated challenge history.

\subsection{Temporal Enforcement}

A single audit does not suffice to establish continuity.
An adversary that disrupts the witness state may still succeed in
isolated rounds.
However, as audits accumulate across multiple rounds, inconsistencies
compound \cite{bellare_reset}.

An adversary attempting to fork the execution must either duplicate the
quantum witness or approximate it classically.
Both strategies inevitably destroy or randomize the encoded phase
information \cite{nocloning,dieks}.
As a result, the probability that multiple forked executions all pass
audits decreases rapidly with the length of the audit window
\cite{helstrom}.

\subsection{Design Rationale}

This construction deliberately separates three roles:
entanglement provides a shared quantum resource,
challenge-dependent phase updates encode temporal information, and
complementary-basis audits reveal inconsistencies caused by disturbance
\cite{ghz,helstrom}.
While the GHZ-based instantiation is simple, it captures the essential
mechanism underlying QSCW.
More sophisticated constructions may improve noise tolerance, reduce
resource requirements, or provide stronger security guarantees, but the
fundamental role of temporal enforcement remains unchanged
\cite{nisq}.

\begin{algorithm}[t]
\caption{GHZ-Based QSCW Protocol}
\label{alg:ghz_protocol}
\begin{algorithmic}[1]
\STATE \textbf{Initialization:} 
\STATE \quad Prover prepares $\ket{\psi_0} = \frac{1}{\sqrt{2}}(\ket{0}^{\otimes n} + \ket{1}^{\otimes n})$.
\STATE \quad History phase $\phi_0 = 0$.

\STATE \textbf{Round $t$ Update:}
\STATE \quad Verifier sends random challenge bits $c_t \in \{0,1\}^k$.
\STATE \quad $p \leftarrow \text{parity}(c_t)$.
\IF{$p == 1$}
    \STATE Prover applies $Z$ gate to first qubit: $\ket{\psi_t} = Z_1 \ket{\psi_{t-1}}$.
    \STATE (Effect: $\ket{\mathrm{GHZ}} \leftrightarrow \ket{\mathrm{GHZ}^-}$)
\ENDIF

\STATE \textbf{Audit:}
\STATE \quad Verifier chooses basis $B \in \{X, Z\}$ randomly.
\STATE \quad Prover measures all qubits in basis $B$, sends outcome $m$.
\STATE \quad \textbf{Verify:} Check if $m$ parity matches accumulated phase history.
\end{algorithmic}
\end{algorithm}

\section{Security Analysis and Discussion}

This section analyzes the security properties of the Quantum State
Continuity Witness (QSCW) and interprets the experimental results in the
context of the Quantum State Continuity Problem (QSCP).
Our analysis focuses on fork resistance, adversarial capabilities, and
the role of temporal enforcement \cite{bellare_reset}.

\subsection{Why Stateless Approaches Fail}

Stateless authentication mechanisms, whether classical or quantum,
evaluate each interaction in isolation.
As a result, they cannot distinguish between a single evolving execution
and multiple forked executions that independently satisfy per-round
checks \cite{attestation,bellare_reset}.
Our experimental results confirm this limitation.
Stateless quantum audits achieve high acceptance rates for honest
behavior but exhibit fork success rates close to random guessing
\cite{helstrom}.
This behavior persists across noise levels, qubit counts, and sampling
parameters.

These observations demonstrate that quantum mechanics alone does not
guarantee continuity.
Without temporal linkage between rounds, quantum evidence can be
approximated or replayed with non-negligible success
\cite{nocloning,helstrom}.

\subsection{Effect of Temporal Enforcement}

Temporal enforcement fundamentally changes the security landscape.
By forcing the witness state to evolve as a function of the interaction
history, QSCW ties future validity to past coherence \cite{nocloning}.
An adversary attempting to fork the execution must disrupt the witness
state at the branching point, either by measurement or approximation.
While such disruption may go undetected in isolated rounds, its effects
accumulate over time.
Each additional audit imposes a new constraint on the set of states
compatible with acceptance \cite{helstrom}.
Our results show that the probability of successful forked acceptance
decreases exponentially with the audit window length.
This exponential decay is consistent with the intuition that each round
reduces the adversary’s remaining degrees of freedom
\cite{bellare_reset}.

\subsection{Adversary Models}

We evaluate QSCW under several adversary models reflecting realistic
system threats.
Memoryless adversaries that replace the witness with classical or
separable approximations are rapidly detected \cite{helstrom}.
Adversaries with limited memory or partial knowledge of past challenges
exhibit increased success rates but remain strongly suppressed as the
audit window grows \cite{bellare_reset}.
In contrast, an ideal coherent adversary capable of maintaining perfect
quantum coherence across time and branches can defeat continuity
enforcement \cite{nocloning,dieks}.
This adversary represents a theoretical upper bound rather than a
realistic deployment threat.
QSCW does not claim security in this regime.

\subsection{Robustness to Noise and Parameters}

A key observation from our evaluation is that fork resistance is largely
insensitive to system parameters.
Varying the number of qubits, measurement shots, and audit thresholds
has minimal impact on the fork success rate.
Noise primarily affects the acceptance rate for honest executions rather
than the adversary’s relative advantage \cite{nisq}.
This robustness suggests that QSCW’s security properties arise from
structural features of temporal enforcement rather than fine-tuned
parameters.
Such behavior is desirable for practical systems, where precise control
over noise and calibration may be infeasible \cite{nisq}.

\subsection{Interpretation and Implications}

The combined results support the central thesis of this work: execution
continuity is a distinct security property that cannot be achieved
through stateless mechanisms alone \cite{bellare_reset}.
QSCW demonstrates that enforcing temporal evolution of quantum evidence
provides a qualitative security improvement against fork attacks.
Importantly, this improvement does not rely on large-scale quantum
resources or precise tuning.
Instead, it emerges from the cumulative effect of simple audits applied
over time \cite{helstrom}.
This observation highlights the potential of temporal enforcement as a
general design principle for quantum-aware security systems
\cite{nisq}.

\section{Evaluation Methodology and Results}

This section describes the experimental methodology used to evaluate
the Quantum State Continuity Witness (QSCW) and presents the resulting
observations.
The evaluation is designed to assess fork resistance, robustness, and
parameter sensitivity under realistic system constraints
\cite{bellare_reset,nisq}.

\subsection{Experimental Setup}

We evaluate both stateless and temporal instantiations of QSCW using
classical simulation of quantum states.
The temporal instantiation follows the construction described in
Section~VI, where a GHZ-based witness evolves over time in response to
verifier challenges \cite{ghz}.
Each experiment proceeds over a fixed number of rounds.
At each round, the verifier selects an audit basis uniformly at random
and checks consistency with the expected witness evolution.
We report two primary metrics:
\begin{itemize}
    \item \emph{Audit Pass Rate (APR)}, defined as the fraction of rounds
    accepted for an honest execution, and
    \item \emph{Fork Success Rate (FSR)}, defined as the probability that
    all branches of a forked execution are accepted within the audit
    window.
\end{itemize}

Unless otherwise stated, experiments are conducted with a fixed number
of qubits, measurement shots, and audit thresholds.
Each data point is obtained by averaging over multiple independent
trials to reduce statistical variance \cite{helstrom}.

\subsection{Baseline: Stateless Auditing}

\begin{algorithm}[H]
\caption{Stateless Baseline (No Temporal Linkage)}
\label{alg:stateless_baseline}
\begin{algorithmic}[1]
\STATE \textbf{For each round $t$:}
\STATE Prover re-initializes a fresh state $\rho^{(fresh)}$ (no memory of $\rho_{t-1}$).
\STATE Prover answers challenge $c_t$ using only the current round state.
\STATE Verifier audits $E_t$ independently per round.
\STATE \textbf{Note:} Since rounds are independent, multiple forks can satisfy per-round audits.
\end{algorithmic}
\end{algorithm}

We first evaluate a stateless quantum audit baseline.
In this setting, the witness is reinitialized at each round and carries
no temporal information \cite{bellare_reset}.
The results show that stateless auditing achieves near-perfect APR for
honest executions.
However, the FSR remains close to random guessing.
Forked executions are therefore able to pass audits independently with
non-negligible probability \cite{helstrom}.
This baseline confirms that quantum measurements alone do not enforce
continuity and motivates the need for temporal linkage between rounds.

\subsection{Temporal QSCW Performance}

We next evaluate the temporal QSCW construction.
In contrast to the stateless baseline, temporal QSCW maintains a single
evolving witness across rounds.
Our results show that temporal QSCW preserves high APR for honest
executions across all tested configurations.
At the same time, FSR decreases sharply as the audit window length
increases.
This behavior is consistent with exponential suppression of fork
success \cite{bellare_reset}.
These findings demonstrate that temporal enforcement introduces a
qualitative separation between honest and forked executions.

\subsection{Audit Window Scaling}

\begin{figure}[H]
    \centering
    \includegraphics[width=0.95\linewidth]{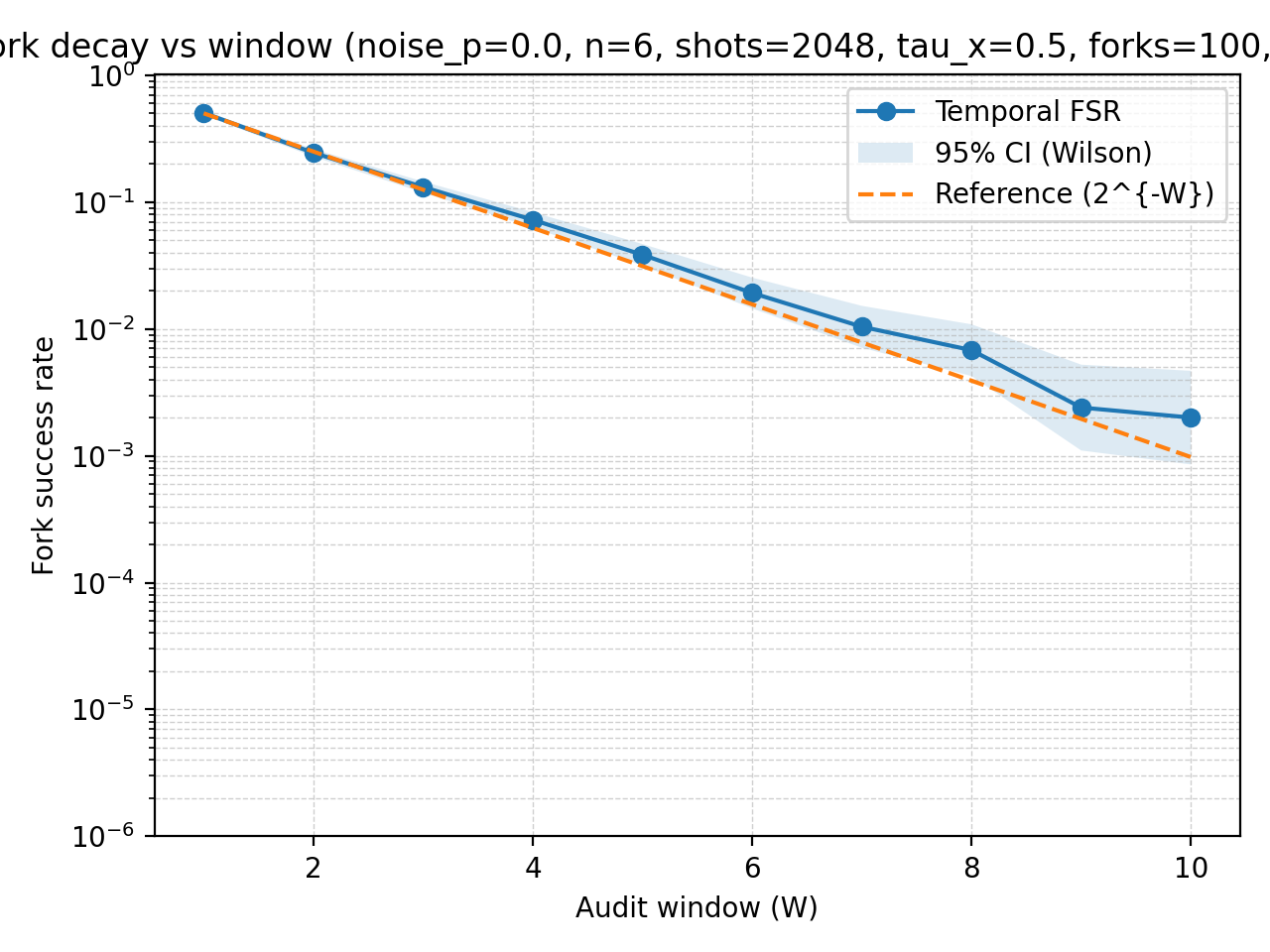}
    \caption{Fork success rate (FSR) vs audit window length $W$ under the temporal QSCW construction.}
    \label{fig:window_sweep}
\end{figure}

To quantify the effect of temporal enforcement, we sweep the audit
window length and measure the resulting FSR.
We observe that the logarithm of FSR decreases approximately linearly
with the window size, indicating exponential decay \cite{bellare_reset}.
This scaling behavior supports the intuition that each additional audit
round further constrains the adversary’s ability to maintain multiple
consistent execution branches.~\ref{fig:window_decay_fit}

\subsection{Robustness to Noise}

We evaluate the impact of noise by introducing controlled perturbations
to the simulated witness state.
As noise increases, the APR for honest executions degrades gradually, as
expected \cite{nisq}.

In contrast, the FSR remains largely unchanged.
Noise primarily affects availability rather than adversarial advantage,
indicating that fork resistance arises from structural temporal
properties rather than fragile parameter choices.

\subsection{Adversary Model Comparison}

Memoryless adversaries are detected quickly.
Limited-memory adversaries exhibit higher success rates but remain
strongly suppressed as the audit window grows.
An ideal coherent adversary defines a theoretical upper bound rather than
a realistic deployment threat~\ref{fig:attacker_k_sweep}
~\ref{fig:window_sweep_models}
\cite{nocloning,dieks}.

\subsection{Summary of Results}

Taken together, the evaluation demonstrates that temporal enforcement
provides a clear security advantage over stateless approaches.
QSCW achieves strong fork resistance without relying on large quantum
resources or precise parameter tuning.
The results support the use of QSCW as a building block for enforcing
execution continuity in quantum-aware security systems \cite{nisq}.

\section{Related Work}

This section situates the Quantum State Continuity Witness (QSCW) within
the broader landscape of authentication, attestation, and
quantum-enhanced security mechanisms.
We emphasize how QSCW differs from prior approaches in both objective
and design \cite{attestation,bellare_reset}.

\subsection{Classical Authentication and Attestation}

Classical device authentication mechanisms rely on cryptographic
primitives such as digital signatures, challenge--response protocols,
and hardware-backed key storage \cite{goldreich}.
Remote attestation frameworks extend this model by verifying software
integrity and configuration state, often using trusted execution
environments or secure enclaves \cite{attestation}.
While these approaches are effective for identity verification, they do
not explicitly enforce execution continuity.
A device that successfully authenticates at multiple points in time may
still represent multiple divergent executions derived from a single
authenticated past \cite{bellare_reset}.
Fork attacks therefore remain largely unaddressed at the authentication
layer.

\subsection{Post-Quantum Cryptography}

Post-quantum cryptography replaces classical cryptographic primitives
with constructions believed to resist quantum attacks \cite{postquantum}.
These techniques preserve the existing authentication and attestation
model while strengthening its computational assumptions.
However, post-quantum cryptography does not alter the structure of the
identity proof itself.
Authentication remains based on key possession and mathematical
hardness, and execution continuity is not enforced \cite{bellare_reset}.
QSCP is orthogonal to post-quantum cryptography and can be applied
independently of the underlying cryptographic primitives.

\subsection{Quantum Authentication and Identification}

Several works have explored quantum-assisted authentication and
identification protocols, including quantum challenge--response schemes
and quantum identification protocols \cite{damgaard_qid}.
These approaches leverage quantum states to prevent replay or
impersonation within a single protocol execution.
Most such schemes are stateless, treating each authentication attempt as
an independent event \cite{helstrom}.
As a result, they do not address fork attacks that span multiple rounds
or extended periods of interaction.
In contrast, QSCW explicitly targets temporal continuity by linking
evidence across rounds \cite{bellare_reset}.

\subsection{Physical Unclonable Functions}

Physical Unclonable Functions (PUFs) and their quantum variants exploit
manufacturing variability or quantum uncertainty to generate
device-specific responses \cite{puf_survey}.
PUFs aim to provide a hardware fingerprint that is difficult to clone or
predict.
Quantum PUF proposals similarly focus on uniqueness and unpredictability
\cite{quantum_puf}.
However, PUF-based approaches are inherently stateless: responses depend
only on the current challenge and physical structure.
They do not encode or enforce temporal evolution across interactions.
QSCW differs fundamentally in objective.
Rather than extracting device-specific randomness, QSCW enforces that a
single execution evolves continuously over time \cite{bellare_reset}.
In this sense, QSCW addresses liveness and continuity rather than
fingerprinting.

\subsection{Quantum Key Distribution and Related Primitives}

Quantum key distribution (QKD) and related primitives leverage the
no-cloning theorem and measurement disturbance to establish secure
communication channels \cite{qkd}.
These protocols demonstrate the power of quantum mechanics to enforce
security guarantees that are impossible classically.
However, QKD treats quantum states as ephemeral resources consumed within
a single session.
It does not enforce continuity of quantum state across extended
interaction periods \cite{nocloning}.
QSCW instead repurposes quantum properties to bind present validity to
past evolution \cite{bellare_reset}.

\subsection{Summary}

Existing work in authentication, attestation, and quantum security
addresses identity verification, secrecy, or randomness extraction
\cite{goldreich,attestation}.
None explicitly target execution continuity as a first-class security
goal.
QSCW complements prior approaches by introducing temporal enforcement as
a primitive capability.
Rather than replacing existing mechanisms, it augments them with a
continuity layer that suppresses fork attacks through cumulative
auditing \cite{bellare_reset}.

\section{Conclusion and Future Work}

This work introduced the Quantum State Continuity Problem (QSCP), a
security objective that is orthogonal to identity authentication and
entity verification \cite{attestation,bellare_reset}.
QSCP focuses on enforcing that a system’s present execution is a
legitimate continuation of a unique past execution, rather than merely
proving possession of valid credentials.
To address QSCP, we proposed the Quantum State Continuity Witness (QSCW)
as a minimal security primitive.
QSCW enforces continuity by requiring quantum evidence to evolve
statefully over time, linking present validity to past interaction
history \cite{nocloning,helstrom}.
We demonstrated through simulation that stateless quantum audits are
insufficient to suppress fork attacks, while temporal enforcement
achieves exponential reduction in fork success probability
\cite{bellare_reset}.
Our evaluation showed that this security improvement is robust across a
wide range of parameters, including noise levels, qubit counts,
measurement shots, and audit thresholds \cite{nisq}.
These results indicate that continuity enforcement arises from structural
features of temporal evolution rather than fragile parameter tuning or
large quantum resources.
Importantly, QSCW is not intended to replace existing authentication or
attestation mechanisms \cite{attestation}.
Instead, it serves as an augmenting building block that provides a
capability absent from current systems: the ability to detect and
suppress forked executions.
By separating execution continuity from identity authentication, QSCP
clarifies a previously underexplored dimension of system security
\cite{bellare_reset}.

This work opens several directions for future research.
First, protocol-level integration of QSCW with existing authentication
frameworks remains an open problem.
Second, stronger constructions may improve noise tolerance or reduce
quantum resource requirements \cite{nisq}.
Third, characterizing the boundary between realistic adversaries and
ideal coherent attackers may yield deeper insight into the limits of
continuity enforcement \cite{nocloning,dieks}.
Finally, experimental realizations on near-term quantum hardware could
inform practical deployment considerations \cite{nisq}.
By framing execution continuity as a first-class security goal and
demonstrating a viable quantum-assisted primitive, this work lays the
foundation for continuity-aware security mechanisms in quantum-aware
systems \cite{attestation,bellare_reset}.

\bibliographystyle{IEEEtran}

\end{document}